\documentclass[preprint,11pt]{elsarticle}

\usepackage{color,xcolor}
\usepackage{nicematrix}
\usepackage{array}
\usepackage{amsmath,amssymb,amsfonts,amsthm,bm}
\usepackage{mathrsfs}
\usepackage{booktabs}
\usepackage{threeparttable}
\usepackage{multirow}
\usepackage{arydshln}

\usepackage{graphicx}

\usepackage{cases}
\usepackage{footnote}
\usepackage[hidelinks,draft=false]{hyperref}
\usepackage[capitalize, nameinlink]{cleveref}
\usepackage[ruled,linesnumbered]{algorithm2e}
\usepackage{textcomp}
\usepackage{url}

\newtheorem{thm}{Theorem}

\newtheorem{rem}{Remark}
\newtheorem{defn}{Definition}

\journal{Applied Energy}

\begin{document}

\begin{frontmatter}



\title{A Resilience-Oriented Centralised-to-Decentralised Framework for Networked Microgrids Management}


\author[inst1]{Pudong Ge}

\affiliation[inst1]{organization={Department of Electrical and Electronic Engineering, Imperial College London},
            city={London},
            postcode={SW7 2AZ}, 
            country={United Kingdom}}

\author[inst1]{Fei Teng}
\author[inst2]{Charalambos Konstantinou}

\affiliation[inst2]{organization={CEMSE Division, King Abdullah University of Science and Technology (KAUST)},
            addressline={Thuwal 23955-6900}, 
            city={Thuwal},
            country={Saudi Arabia}}

\author[inst3]{Shiyan Hu}

\affiliation[inst3]{organization={School of Electronics and Computer Science, University of Southampton},
            city={Southampton},
            postcode={SO17 1BJ}, 
            country={United Kingdom}}

\begin{abstract}
This paper proposes a cyber-physical cooperative mitigation framework to enhance power systems resilience {against power outages caused by extreme events, e.g., earthquakes and hurricanes}. Extreme events can simultaneously damage the physical-layer electric power infrastructure and the cyber-layer communication facilities. 
Microgrid~(MG) has been widely recognised as an effective physical-layer response to such events, however, the mitigation strategy in the cyber lay is yet to be fully investigated. Therefore, this paper proposes a resilience-oriented centralised-to-decentralised framework to maintain the power supply of critical loads such as hospitals, data centers, etc., under extreme events.
For the resilient control, controller-to-controller~(C2C) wireless network is utilised to form the emergency regional communication when centralised base station being compromised. Owing to the limited reliable bandwidth that reserved as a backup, the inevitable delays are dynamically minimised and used to guide the design of a discrete-time distributed control algorithm to maintain post-event power supply. The effectiveness of the cooperative cyber-physical mitigation framework is demonstrated through extensive simulations in MATLAB/Simulink.
\end{abstract}

\begin{graphicalabstract}
\includegraphics[width=1.0\textwidth]{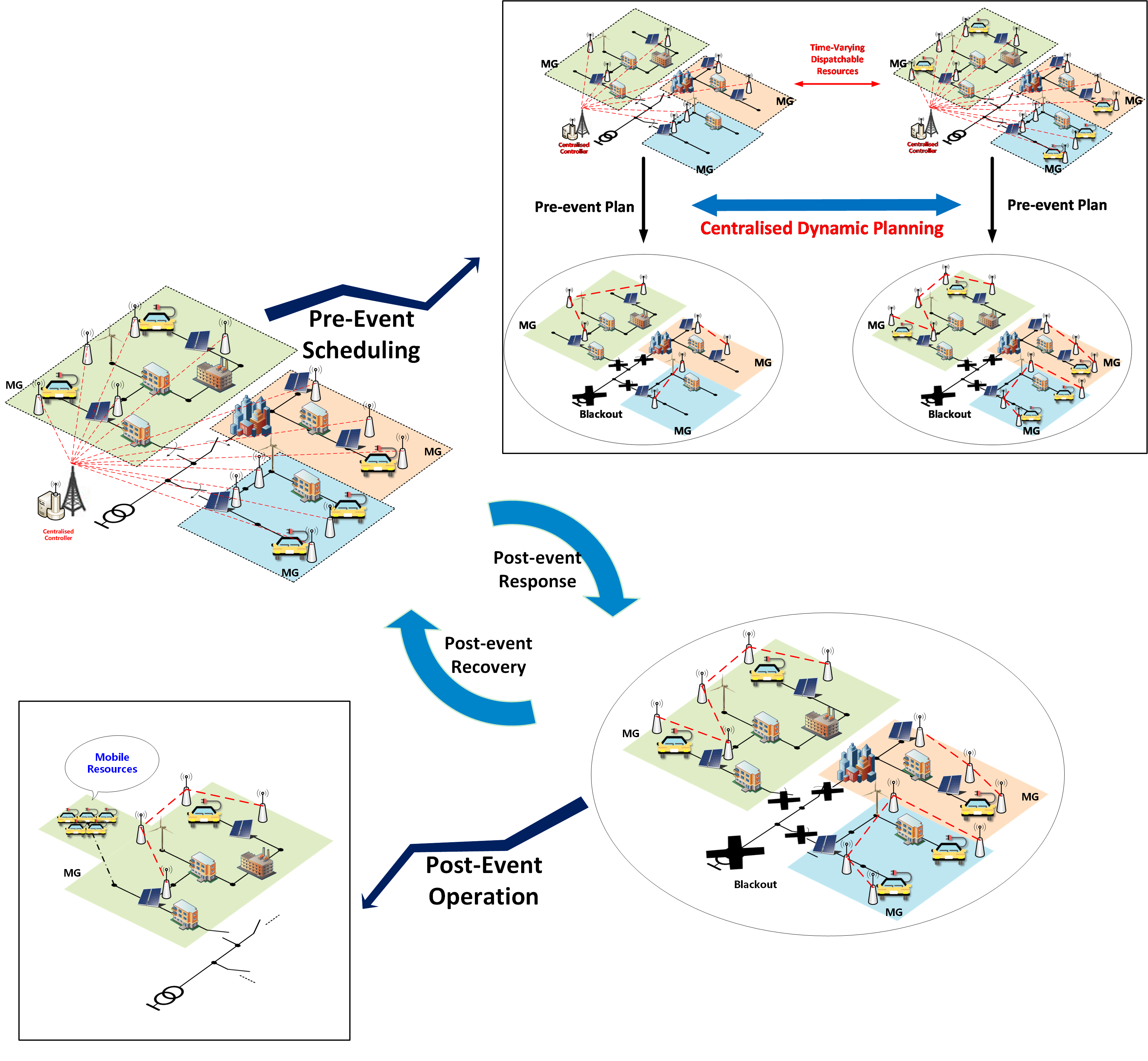}
\end{graphicalabstract}

\begin{highlights}
\item A centralised-to-decentralised framework is proposed to tackle adverse cyber-physical effects (ACPEs) under extreme events. 
\item In the cyber layer, a dynamic resource allocation, over a limited wireless bandwidth, effectively and quantitatively optimises the non-negligible delays and informs the design of the control algorithm.
\item A discrete-time distributed control system based on optimised distributed communication network simplifies the selection of control gains.
\end{highlights}


\end{frontmatter}


\section{Introduction}\label{sec:intro}
Global warming drives the energy supply transition from traditional fossil fuel based power generation to renewable energy resources. This transition has been widely recognised as one of the most significant developing pathways promoting low/zero-carbon societies~\cite{creutzig2014catching}. Rapidly developing renewable energy generators gradually dominate power systems especially in distribution power networks~\cite{apostolopoulou2016interface,wang2020sustainable}. During the energy transition process, existing climate change leads to an increase in the frequency, intensity and duration of severe weather events~\cite{li2017networked,panteli2015influence}. Extreme weather conditions pose huge threats to power supply infrastructures, thereby leading to undesired power interruptions or blackouts. Hence, the concept of resilience under extreme events has been gradually recognised as a key requirement for future energy systems~\cite{panteli2015influence,wang2016research, konstantinou2021towards}. 

Effectively utilising renewable distributed generators (DGs) to provide emergency power supply for critical loads in the form of microgrid~(MG) is a widely used solution to enhance the resilience of power supply~\cite{wang2020sustainable,olivares2014trends}. For example, the Consortium for Electric Reliability Technology Solutions (CERTS)-enabled MG maintained power, water, heat of the Brevoort building in Greenwich Village, NY, USA, during the week of wide spread utility outages due to Hurricane Sandy in late 2012~\cite{panora2014real}. In Japan, Sendai MG and Roppongi Hills MG demonstrated that well-developed localised energy systems are essential to handle critical emergency resulting from earthquakes and tsunamis~\cite{marnay2015japan}.

Currently, the research on regional MGs providing post-event power supply mostly focuses on the control strategy after islanding operation. The control methods of islanded MGs, either centralised or decentralised, have been widely investigated to regulate the frequency and voltage in the presence of renewable energy generators~\cite{khayat2019secondary,ge2021event,ge2020resilient}. 
In addition, the concept of dynamic MGs, including reconfigurable cyber and physical layers, has been proposed to enable the autonomous operation of distribution systems~\cite{du2020dynamic}, but the cyber solution and cyber-physical coordination, in the event of simultaneous cyber and physical damage, has not been clarified.

On the other hand, wireless communication technologies, e.g., the fifth-generation~(5G), have been widely investigated to support the efficient operation and coordination of massive distributed energy resources (DERs)~\cite{parikh2010opportunities}. Leveraging advanced communication technologies, the operation of power systems is becoming intelligent towards a highly cyber-physical fusion~\cite{yu2016smart}. Although advanced communication technologies enable the real-time efficient centralised control framework, which is superior to the decentralised one in terms of control performance and implementation efficiency, such centralised framework suffers from a single-point failure. In addition, base stations that support the management of wireless resources are vulnerable to natural disasters or cyber-attacks, and they may fail to function if losing the backhaul connection to the core network or being physically damaged~\cite{alqahtani2018disaster, 9351954}. Hence, for instance, ad-hoc communication technology has been utilised to realise the self-organised MGs in response to disasters~\cite{alqahtani2018disaster,deepak2019overview}. However, further detailed cyber-layer scheduling and implementation in response to contingencies have not been investigated.

Considering the limited occurrence of extreme events, the existing centralised control framework enabled by 5G networks should be fully utilised because of its advantages in achieving the global economical efficiency and easy integration into the existing centralised control framework (SCADA system). However, the vulnerability  of centralised framework against single-point failure should be improved. The promising solution is to design a transition scheme from centralised framework to decentralised framework utilising the power electronic devices and the wireless communication technologies in response to extreme conditions. The cyber-physical collaborative transition and response strategy during the pre-event and post-event periods, especially the cooperative design of communication network and control strategy, has not been examined in the literature. In this paper, we design a resilience-oriented centralised-to-decentralised framework of networked MGs to maintain the critical power supply by utilising MG clusters isolated from the distribution system, thus enhancing the power supply resilience. Under such centralised-to-decentralised transition, the communication network is converted from base station supporting mode under normal operating conditions to controller-to-controller~(C2C) mode under extreme conditions. The C2C communication only requires wireless module equipped at the local controller, which is originally necessary for receiving the instructions in the pre-event normal condition. To summarise, the contributions of the paper are listed as follow:
\begin{enumerate}
    \item A centralised-to-decentralised framework is proposed to tackle adverse cyber-physical effects (ACPEs), which can simultaneously benefit from the efficiency of centralised framework (centralised controller and centralised communication) in normal operations and the resilience of decentralised framework (distributed controller and C2C communication) under extreme events.
    \item In the cyber layer, a dynamic resource allocation based C2C communication protocol, over a limited wireless bandwidth, is proposed to facilitate emergency communication under extreme events. The communication resource allocation model effectively and quantitatively optimises the non-negligible delays and informs the design of the control algorithm.
    \item A discrete-time distributed control system is co-designed along with the dynamically-scheduled wireless network solution. A delay-dependent sampling interval is proposed based on optimised communication resources, which simplifies the selection of control gains and enables the plug-and-play operation of MGs during the post-event period.
\end{enumerate}

The remainder of this paper is organised as follows: \cref{sec:frame} introduces the detailed framework, while \cref{sec:cyber_schedule} and \cref{sec:control} provide the system design method from cyber layer and physical layer respectively. In \cref{sec:results}, simulation results are given, and \cref{sec:conclusion} concludes the paper.

\section{Centralised-to-Decentralised Resilient Framework: A Cyber-Physical Perspective} \label{sec:frame}

The resilient response to contingencies from a cyber-physical perspective consists of three scenarios: cyber contingencies, physical contingencies, cyber-physical contingencies. 
The research in this paper focuses on {resilience enhancement against failures or damages in both cyber and physical layers. For instance, natural disasters, e.g., earthquakes, hurricanes, and flooding, can destroy both power and communication infrastructures, and cyber-attacks can cause cascading failures leading to both power line damage and unreliable communication.} Inspired by~\cite{konstantinou2021resilient}, we can define such events as the following:
\begin{defn}
    \textbf{Adverse Cyber-Physical Effects (ACPEs)} involve single or the combination of the following extreme events, e.g., natural disasters such as hurricanes, earthquakes, wildfires, ‘silent errors’ due to components and manufacturing variability failures, hardware or software faults of smart monitoring devices due to bugs in the code (e.g., operating system, compilers, libraries, etc.), natural effects such as bit flips induced by hardware failures, drive failures, cosmic rays, cyber-attacks, or even faults involving the infrastructure design and implementation. ACPEs drastically affect the results of cyber-physical algorithms in power systems, and subsequently the operations of both cyber and physical components deployed in critical infrastructures.
\end{defn}

\subsection{Overview of Centralised-to-Decentralised Framework}
\begin{figure}[!ht]
    \centering
    \includegraphics[width=1.0\textwidth]{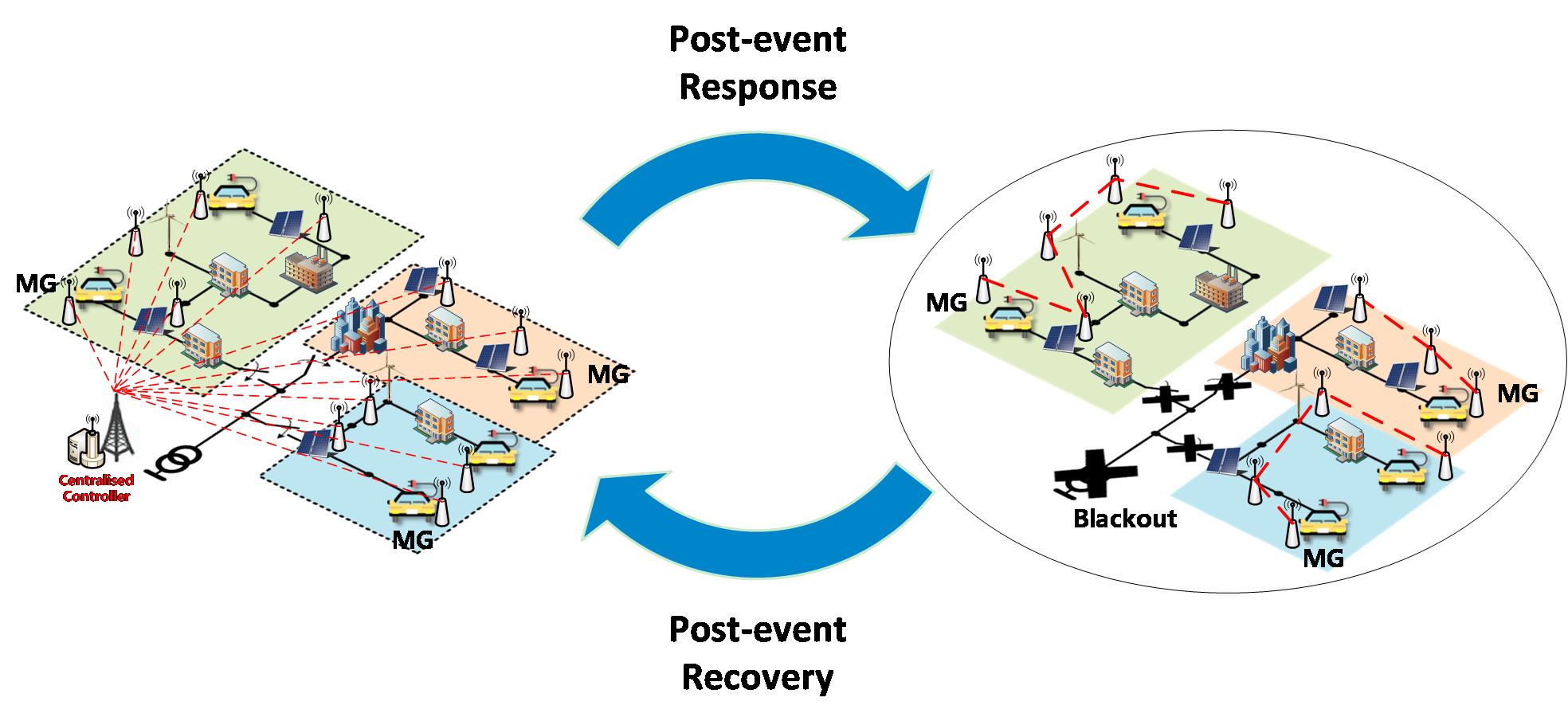}
    \caption{Blackout response framework -- a centralised-to-decentralised method.}
    \label{fig:C2D}
\end{figure}

A resilience-oriented centralised-to-decentralised framework, as in \cref{fig:C2D}, is proposed in response to ACPEs, which trigger the islanding operation in the physical layer and transition of the control and communication mode from the centralised to the decentralised. The centralised control structure under normal operation is served by a centralised controller, normally implemented in the substation. Such centralised controller coordinates all dispatchable resources (e.g., DGs, controllable loads, electric vehicles) to maintain the grid frequency, voltage stability, and economical dispatch through using wired or wireless communication, {e.g., the 5G network{\footnote{Compared to wired communication, wireless communication is more affordable to coordinate massive distributed generators. Among wireless technologies, owing to massive distributed generators needed to be regulated, 5G with its abundant derivative technology~\cite{chettri2019comprehensive} is a promising solution thanks to its high bandwidth and wide coverage.}}~\cite{wu2017overview,feng2012survey}} supported by base stations. On the other hand, the decentralised structure under extreme events, whose priority is safety, is the emergency response to maintain the critical energy supply as much as possible by utilising available localised distributed resources and grid-forming techniques through device-to-device~(D2D) ad-hoc wireless communication. Such a centralised-to-decentralised framework can complement existing centralised control structures, and benefit from the efficient and flexible grid formation of the decentralised structure. The detailed cyber-physical solution is outlined in \cref{tab:cps_solution}.

\begin{table}[!ht]
\centering
\caption{Cyber-physical solution of centralised-to-decentralised framework.}
\label{tab:cps_solution}
\begin{tabular}{lll}
\hline
 & Normal condition  & Extreme condition \\ \hline
Physical solution & Main grid and DERs & DER-based MGs \\
\multirow{2}{*}{Cyber solution} & Wired/wireless & D2D-enabled wireless \\
& Centralised coordination & Decentralised/distributed \\
Objectives & Economical operation & Critical power supply  \\
Priority & Optimality & Safety \\
\hline
\end{tabular}
\end{table}

\subsection{Centralised-to-Decentralised Post-Event Response Process}
\begin{figure}[!ht]
    \centering
    \includegraphics[width=0.8\textwidth]{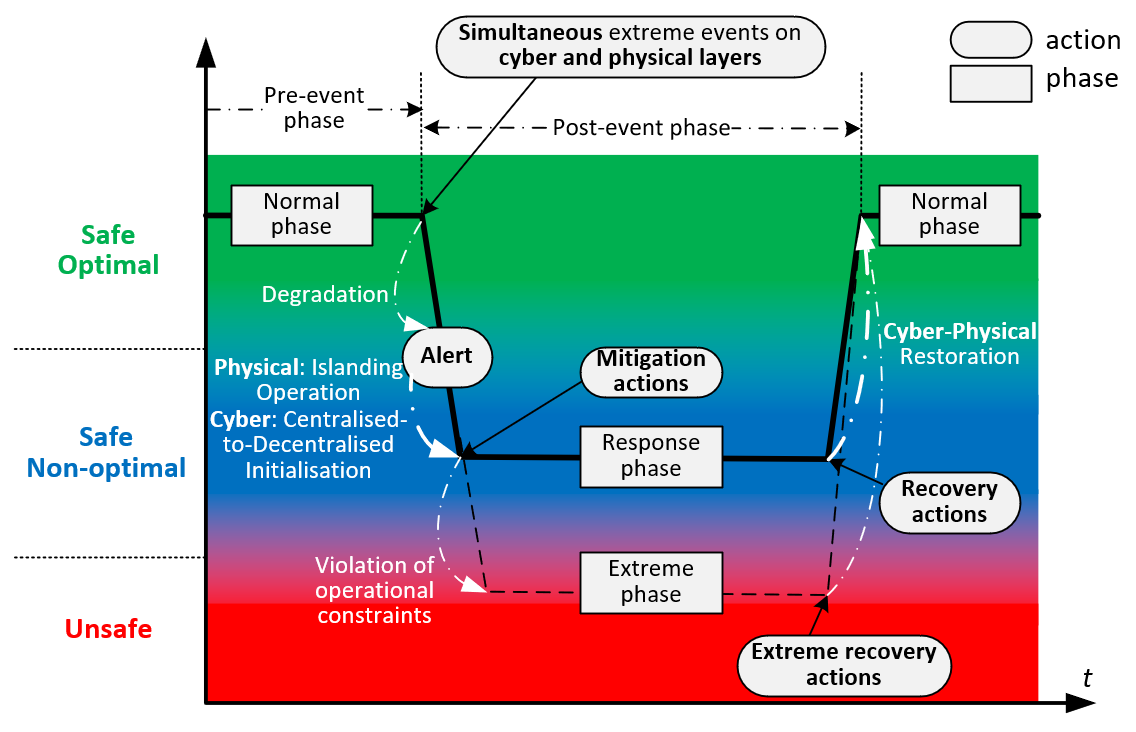}
    \caption{Time-sequential diagram of the centralised-to-decentralised framework.}
    \label{fig:time_response}
\end{figure}

The centralised-to-decentralised transition functions as a post-event response against ACPEs. To form such transition, a dynamic cyber-physical scheduling scheme is required to clearly guide the immediate decentralised and localised energy supply. Under normal operation, the post-event cyber-physical response schedule is dynamically optimised in the central controller and sent, together with other control signals, to the local controllers. The sequential diagram of the cyber-physical collaborative response framework is outlined in \cref{fig:time_response}, and the centralised-to-decentralised framework focuses on the transition to the ``response phase" using the proposed mitigation actions in this paper.

Once damage or failure occurs after extreme events, the cyber-physical solution for power networks operating in the normal condition could be out of service. Physically, the main grid support is lost, which can be detected by the control and protection system using electrical state sensing. The successful detection triggers breakers switching off to enable networked MGs being split into islanded MGs. {From a cyber perspective, if the local controller loses the connection to the centralised controller for a period of time (a threshold value), an emergency wireless network formation is triggered to form as the pre-scheduling.} The controllers that enable direct D2D communication can move to the emergency mode to form an ad-hoc and self-organised emergency wireless network serving for a regional islanded MG. A reconfiguration protocol of the cyber layer can support the discovery of neighbour nodes~\cite{alqahtani2018disaster,chen2016neighbor} can be adopted or modified. The basic idea is to perform a handshake, i.e., scan and respond to a beacon signal from other nodes, or emit a beacon signal, wait for response, and then utilise the pre-scheduled frequency assignment to establish a communication link.

Hence, during the extreme conditions, each MG maintains the critical energy supply by limited energy capability under emergency scenarios and operates as that scheduled before in both cyber and physical layers. Followed by gradually repaired power supply infrastructure, the operation of networked MGs will recover to the normal condition. It should be noted that the recovery process is out of the scope of this paper.

{
\begin{rem}
    The time and duration of the blackouts caused by extreme events vary case-by-case, from a short time of cyber-attack to a long time of hurricane for instance. The proposed framework can cope with the transition from centralised control to decentralised control when the blackout and cyber infrastructure damages occur simultaneously for both short-time and long-time disasters. However, the short-term and long-term responses require different energy capabilities and long-term response strategies, which are determined and optimised by the installation of renewable energy resources and portable storage connection design. The long-term response is the next step of our centralised-to-decentralised transition framework and is not inside the scope of this paper. We will consider the long-term lasting post-event restoration in our future work.
\end{rem}
}

\section{Dynamic Cyber-Layer Scheduling Considering Time-Varying Dispatchable Resources}\label{sec:cyber_schedule}
\begin{figure}[!ht]
    \centering
    \includegraphics[width=1.0\textwidth]{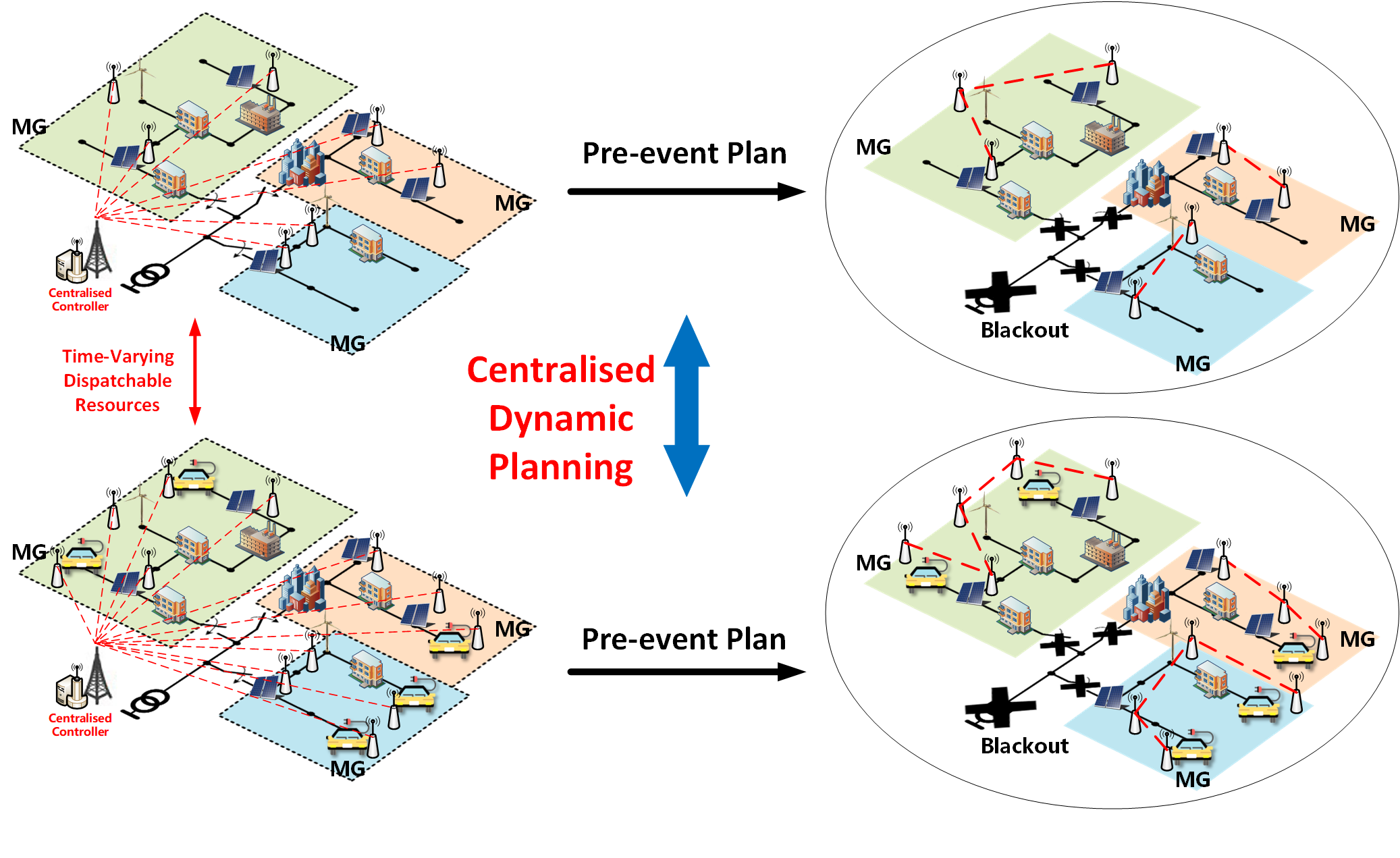}
    \caption{Centralised pre-event dynamic cyber-physical scheduling scheme.}
    \label{fig:pre-event_scheduling}
\end{figure}

{
The dynamic cyber-physical response scheduling is performed during the pre-event stage in the normal condition. The objective of the scheduling is to design cyber-physical solutions in advance in order to accomplish seamless transformation once ACPEs occur. Owing to plug-and-play characteristic of EVs and time-varying availability of dispatchable loads, the scheduling needs to be dynamically optimised under different scenarios. The cyber-layer design utilises distributed C2C wireless communication in virtue of promising D2D techniques~\cite{deepak2019overview,asadi2014survey}, which efficiently benefits from flexible networking mode.}
The dynamic cyber-physical scheduling scheme, as shown in \cref{fig:pre-event_scheduling} divides distributed power systems into regionally localised autonomous MGs according to line breakers.

In each region, cyber-layer solutions are designed independently in terms of bandwidth allocation. In other words, the total backup bandwidth that is reserved for emergency use can be reused because different regions have diverse physically geographic locations, and thus communication inter-regional collisions are assumed to be negligible~\cite{kong2020multicell}.
On the other hand, the mitigation of intra-regional communication collisions is inspired by frequency division duplexing~(FDD) techniques, which utilise frequency separation multiplexing technology to separate the transmitted and received signals. Owing to the two-way data transmission among distributed control in the post-event MGs, FDD can decrease co-channel interference of two-way C2C communication, and hence avoid collisions in the intra-regional area~\cite{wu2017overview}. 

In the remainder of this section, the details of dynamic cyber-physical scheduling are introduced with the objective of quantitatively determining the delay in each intra-regional wireless network, which will guide the control strategy design in \cref{sec:control}.

\subsection{Wireless Network Model}
The controllable DERs in networked MGs vary in different centralised optimisations, where wireless network of DGs that available for emergency use can be modelled by a dynamic undirected graph $\mathcal{G}=\{\mathcal{V,E,A}\}$, where $V=|\mathcal{V}|$ denotes the numbers of available emergency resources. Such graph $\mathcal{G}$ is certainly not connected because of switched-off breakers but contains $\phi=|\Phi|$ connected components representing MGs:
\begin{align}
    \begin{aligned}
        & \mathcal{G} = \bigcup_{\mu=1}^{\phi}{\mathcal{G}_{\mu}} \\ 
        & \mathcal{G}_{\mu}\cup\mathcal{G}_{\nu} = 0,\ \mu\neq\nu,\ \forall \mu,\nu\in\Phi \label{eq:graph}
    \end{aligned}
\end{align}
where $\mathcal{G}_{\mu}=\{\mathcal{V}_{\mu},\mathcal{E}_{\mu},\mathcal{A}_{\mu}\}$ denotes a connected component representing one emergency MG with $V_{\mu}=|\mathcal{V}_{\mu}|$ DGs. \cref{eq:graph} is equivalent to $\sum_{\mu=1}^{\phi}{V_{\mu}}=V$, and apparently adjacent matrix $\mathcal{A}$ has a block diagonalized form
\begin{align}
    \mathcal{A} = \mathrm{blockdiag}{\{\mathcal{A}_{1},\mathcal{A}_{2},\dots,\mathcal{A}_{\phi}\}}
\end{align}
owing to none available inter-regional communication links. 

For the sake of generality and convenience, wireless network $\mathcal{G}_{\mu}$ is discussed in details, and the combination of multi-graph~($\bigcup_{\mu=1}^{\phi}{\mathcal{G}_{\mu}}$) optimisation forms the dynamic cyber-layer scheduling in the centralised controller during the pre-event period.

For wireless network $\mathcal{G}_{\mu}$ modelling the $\mu_{\mathrm{th}}$ autonomous MG, among $V_{\mu}$ DGs, the communication connection is described by a binary matrix $\mathcal{A}_{\mu}=[a^{\mu}_{ij}]\in\mathbb{R}^{V_{\mu}\times V_{\mu}}$, where all elements are $0$ except for $a^{\mu}_{ij}=1,i\neq j$ only if node $j$ has access to data of node $i$, i.e. $(i,j)\in\mathcal{E}_{\mu}$. Through applying FDD technique modelled by undirected graph, there always exists $\mathcal{A}_{\mu}=\mathcal{A}_{\mu}^{T}$. In other words, any $a^{\mu}_{ij}=1,i\neq j$ means two-way communication between nodes $i,j$. In addition, the component $\mathcal{G}_{\mu}$ has the undirected characteristic and connectivity~\cite{hu2021cost,rafiee2010optimal}, leading to
\begin{align}
    \lambda_{2}(\mathcal{L}_{\mu})>0
\end{align}
where $\lambda_{2}(\mathcal{L}_{\mu})$ denotes the  second smallest eigenvalue of the Laplacian of a graph. This constraint can be relaxed using primal–dual variables and has been investigated in ~\cite{rafiee2010optimal}, so we omitted its discussion in this paper.

The backup bandwidth dedicated to emergency use is divided into $L$ sub-carriers in the set $\mathcal{L}=\{1,2,\dots,L\}$. Let $g_{m,l}$ and $p_{m,l}$ be the channel gain and the transmit power for one C2C link $m=(i,j)\in\mathcal{E}_{\mu}$ on the sub-carrier $l\in\mathcal{L}$ respectively. Then, signal-to-noise~(SNR) for C2C link $m$ on the sub-carrier $l$~\cite{goldsmith2005wireless} is expressed as
\begin{align}
    \gamma_{m,l}=\frac{p_{m,l}g_{m,l}}{\sigma^2}\label{eq:SNR}
\end{align}
where $\sigma^2$ denotes the power of additive white Gaussian noise. The channel gain $g_{m}$ of the C2C link $m=(i,j)$ is negatively related to the transmitting distance with fading effects~\cite{liang2017resource}. Combining both path loss and log-normal shadowing, the channel gain is simply modeled as
\begin{align}
    g_{m}=hd_{i,j}^{-\alpha} \label{eq:wireless_gain}
\end{align}
where $h$ is the loss factor combining total fading effects; $d_{i,j}$ is the transmitting distance between the communication link $m=(i,j)$; $\alpha$ is the pathloss exponent.

The reachable instantaneous C2C data rates in bits per second~(bps) are computed through the well-known Shannon formula 
\begin{align}
    R_{m}=w\sum_{l=1}^{L}\beta_{m,l}\mathrm{log}_{2}(1+\gamma_{m,l})\label{eq:shannon_formula}
\end{align}
where $\beta_{m,l}$ is a binary variable, and $\beta_{m,l}=1$ if sub-carrier $l$ is allocated to the C2C link; otherwise $\beta_{m,l}=0$. $w$ denotes the bandwidth of each sub-carrier.

In the emergency wireless network, normal Quality-of-Service~(QoS) requirement is discarded in the constraints, and is reflected by transmission delay $\tau_{m}$
\begin{align}
    \tau_{m}=\frac{L_{\mathrm{packet}}}{R_{m}}\label{eq:delay}
\end{align}
where $L_{\mathrm{packet}}$ denotes the packet size of data transmissions.

In addition, each controller of node $i$ equipped with wireless communication module has its power limitation, which consists of constant circuit power and total transmit power
\begin{align}
    p_{i,\mathrm{total}} = p_{i,\mathrm{cst}} + \sum_{m\in\mathcal{E}_{\mu,i}}\sum_{l=1}^{L}{p_{m,l}} \leq p_{i,\max},\ \forall j\in\mathcal{V}_{i}
\end{align}
where $\mathcal{E}_{\mu,i}$ denotes the edge/link set where node $i\in\mathcal{V}_{\mu}$ shares local information.

\subsection{Service-Oriented Resource Allocation Problem}

Inspired by the best-effort service of wireless network, the optimisation aims to minimise the total communication delay of wireless sub-networks $\{\mathcal{G}_{1},\mathcal{G}_{2},\dots,\mathcal{G}_{\phi}\}$.
Based on the analysis above, the optimisation problem can be formulated as
\begin{subequations}
    \begin{align}
        & \min_{\mathcal{A},\bm{\beta,p}}\quad \sum_{\mu=1}^{\phi}{\big(\tau_{\max}^{\mu}\big)} \tag {9} \label{eq:ra_obj}\\
        & \mathrm{subject\ to:} \notag \\
        & \mathrm{C_{1}:}\quad\mathcal{A}\in\{0,1\}^{V\times V} \label{eq:ra_c1}\\
        & \mathrm{C_{2}:}\quad\mathcal{A}=\mathcal{A}^{T} \label{eq:ra_c2}\\
        & \mathrm{C_{3}:}\quad\mathrm{tr}(\mathcal{A})=0 \label{eq:ra_c3}\\
        & \mathrm{C_{4}:}\quad \lambda_{2}(\mathcal{L}_{\mu})>0,\forall \mu\in\Phi \label{eq:ra_c4}\\
        & \mathrm{C_{5}:}\quad \beta_{m,l}\in\{0,1\},\forall m\in\mathcal{E},l\in\mathcal{L} \label{eq:ra_c5}\\
        & \mathrm{C_{6}:}\quad \sum_{m\in\mathcal{E}_{\mu}}\beta_{m,l}\leq 1,\forall l\in\mathcal{L}, \mu\in\Phi \label{eq:ra_c6}\\
        & \mathrm{C_{7}:}\quad p_{i,\mathrm{cst}} + \sum_{m\in\mathcal{E}_{\mu,i}}\sum_{l=1}^{L}{p_{m,l}} \leq p_{i,\max},\forall i\in\mathcal{V}_{i},\mu\in\Phi \label{eq:ra_c7} \\
        & \mathrm{C_{8}-C_{11}:}\quad \mathrm{Eqs.}~\eqref{eq:SNR}\eqref{eq:wireless_gain}\eqref{eq:shannon_formula}\eqref{eq:delay},\forall \mathcal{G}_{\mu},\mu\in\Phi \label{eq:ra_c8-11}
    \end{align}
\end{subequations}
where $\tau_{\max}^{\mu}=\max_{m\in\mathcal{E}_{\mu}}{\tau_{m}^{\mu}},\forall{\mu}\in\Phi$, $\bm{\beta}=[\beta_{m,l}]\in\{0,1\}^{E\times L}$, and $\bm{p}=[p_{m,l}]\in\mathbb{R}_{\geq0}^{E\times L}$. Eqs.~\eqref{eq:ra_c1} -- \eqref{eq:ra_c4} are constraints derived from the undirected and connected sub-graphs. \cref{eq:ra_c5} and \cref{eq:ra_c6} are the exclusive sub-carrier allocation in the communication links. \cref{eq:ra_c7} represents the wireless module in each DG controller should satisfy the maximum power consumption requirement. Constraints \cref{eq:ra_c8-11} are basic resource allocation equations being a bridge among channels, power consumption and transmitting delay.
The resource allocation optimisation of dynamic cyber-layer scheduling for emergency wireless network expressed by Eqs.~\eqref{eq:ra_obj} is a mixed integer nonlinear optimisation programming (MINLP). {Due to inefficient solving of MINLP, and inspired by~\cite{termehchi2021joint}, we reduce the complex problem into two allocation sub-problems: sub-carrier allocation and power allocation.}

Sub-carrier allocation sub-problem is optimised under a given power allocation:
\begin{align}
    \begin{aligned}
        & \min_{\mathcal{A},\bm{\beta}}\quad \sum_{\mu=1}^{\phi}{\big(\tau_{\max}^{\mu}\big)}\\
        & \mathrm{subject\ to:\ C_{1}-C_{11}}
    \end{aligned}\label{eq:sub-carrier_allocation}
\end{align}
and power allocation sub-problem is optimised under a given wireless network and corresponding sub-carrier assignment:
\begin{align}
    \begin{aligned}
        & \min_{\bm{p}}\quad \sum_{\mu=1}^{\phi}{\big(\tau_{\max}^{\mu}\big)}\\
        & \mathrm{subject\ to:\ C_{7}-C_{11}}
    \end{aligned}\label{eq:power_allocation}
\end{align}
Both problems \cref{eq:sub-carrier_allocation} and \cref{eq:power_allocation} are convex by utilising exponential cones for \cref{eq:shannon_formula} and rotated quadratic cones for \cref{eq:delay} specially. We solve the problem \eqref{eq:ra_obj} by iteratively solving sub-problems \eqref{eq:sub-carrier_allocation} and \eqref{eq:power_allocation} using MATLAB/YALMIP with MOSEK~\cite{lofberg2004yalmip,mosek}. The overall algorithm of proposed resource allocation problem is detailed in Algorithm~\ref{alg:1}. 

\begin{algorithm}[!ht]
    \label{alg:1}
    \LinesNumbered
    \caption{Overall Procedure of Resource Allocation for Dynamic Cyber-Layer Scheduling in Centralised Controller}
    \KwIn{$w$:~sub-carrier~bandwidth; $L$:~number~of~sub-carriers; $\sigma^{2}$:~power~of~additive~noise; $g_{m}$:~channel~gain~of~links; $V_{\mu}$:~number~of~sub-network~nodes;  $L_{\mathrm{packet}}$:~packet~size; $P_{i,\mathrm{cst}}$:~constant~power~of~nodes; $P_{i,\max}$:~maximum~power~of~nodes;}
    \KwOut{$\mathcal{A}$:~scheduled~wireless~network; $\bm{\tau}_{\max}=[\tau_{\max}^{\mu}]\in\mathbb{R}^{\phi}$:~maximum~communication~delay;}
    \BlankLine
    \textbf{Initialisation:} maximum iterations $k_{\max}$; iteration index $k=0$; convergence error $\epsilon$; transmitting power of links on channels~$\bm{p}$\;
    \While{$k<k_{\max}$}{
        \Repeat{$\mathcal{A}(k)-\mathcal{A}(k-1)=\mathbf{0}$ and $\|\bm{\tau}_{\max}(k)-\bm{\tau}_{\max}(k-1)\|_{\infty}\leq\epsilon$}{
        solve \eqref{eq:sub-carrier_allocation} using MATLAB/YALMIP with MOSEK\;
        update wireless network matrix $\mathcal{A}$ and sub-carrier assignment matrix $\bm{\beta}$\;
        solve \eqref{eq:power_allocation} using MATLAB/YALMIP with MOSEK\;
        update transmitting power of links on channels~$\bm{p}$ and communication delay $\bm{\tau}_{\max}$\;
        }
    }
\end{algorithm}

\begin{rem}
    The optimised communication delay $\tau_{\max}^{\mu}$ considered in this paper corresponds to the transmission delay, which defines the time taken to push the packet bits onto the scheduled channel. As it is well known, the communication delay typically consists of transmission delay, propagation delay, processing delay, and queuing delay. In our context, due to limited bandwidth capacity existing in the emergency C2C communication, transmission delay dominates the communication delay, which is why others are omitted here.
    \label{rem:delay}
\end{rem}

\section{Post-Event Response Based on Scheduled Cyber Network and Distributed Consensus Protocol} \label{sec:control}
\begin{figure}[!ht]
    \centering
    \includegraphics[width=1.0\textwidth]{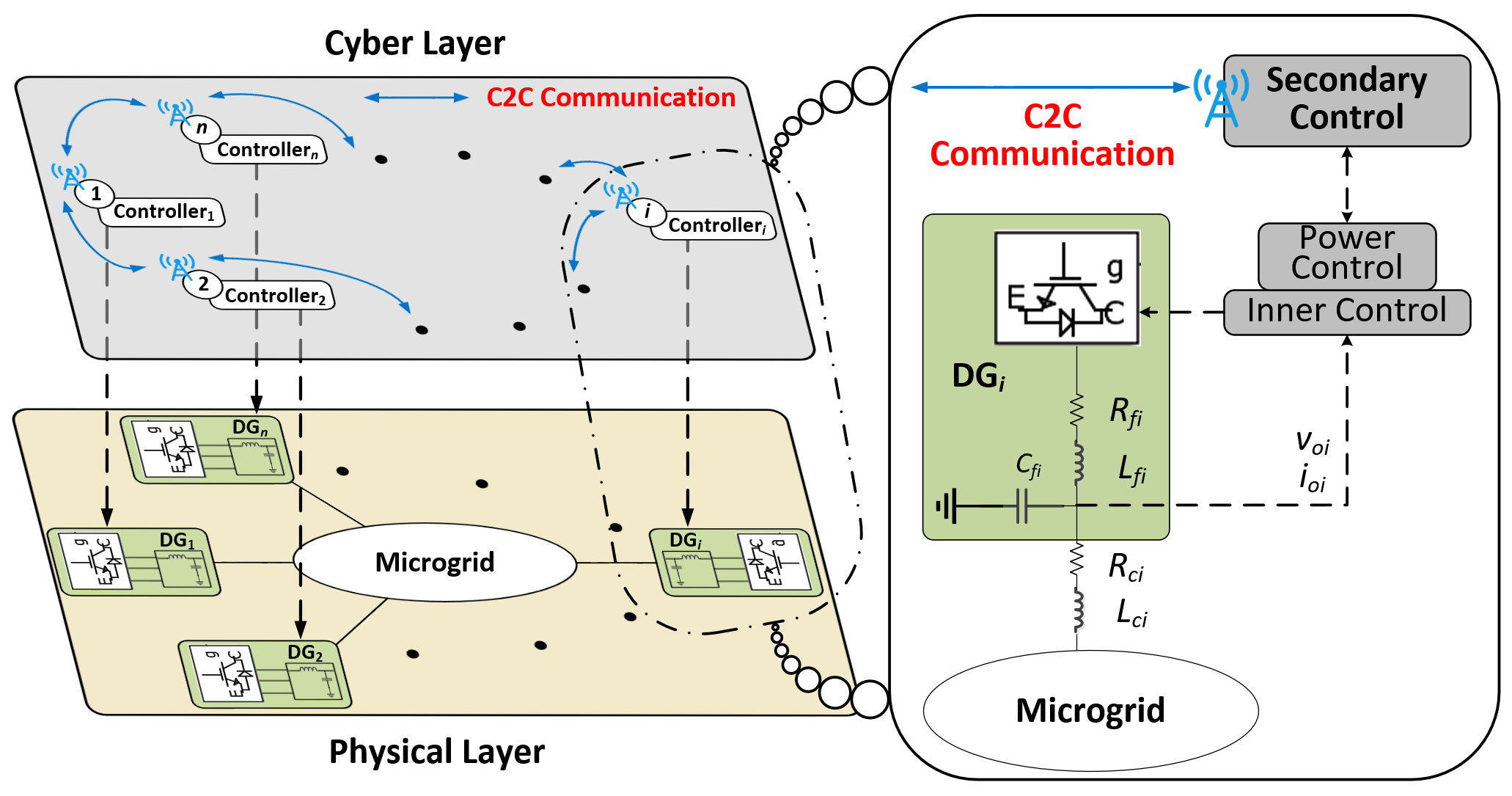}
    \caption{Diagram of a cyber-physical MG and control loops.}
    \label{fig:cps_mg}
\end{figure}

The operation of emergency MG clusters employs the wireless network that is scheduled by the centralised controller as a cyber solution. The physical solution enabling such post-event response utilises available localised energy sources to maintain the critical power supply. More specifically, inverter-based DGs using grid-forming techniques are considered due to its autonomous operation ability. The integrated cyber-physical modelling and structure of such MGs are detailed in \cref{fig:cps_mg}.

\subsection{Physical-Layer Model and Post-Event Response Objectives}
The modelling of emergency MGs focuses on the generators only because the details of temporarily operating MGs are not realistic, especially for those enabling plug-and-play operations. Hence, models focus on the inverter-based DG itself by appropriate approximation. The output active power of DG~$i$ can be expressed as the following with a line impedance $Z_{ij}=R_{ij}+jX_{ij}$ between DG~$i$ and DG~$j$~\cite{simpson2015secondary}:
\begin{align}
    \begin{aligned}
        & P_{i} = P_{i,Load}+\sum_{j=1}^{N_i}\frac{U_{i}U_{j}}{X_{ij}}\sin{(\theta_{i}-\theta_{j})} \\
        & Q_{i} = Q_{i,Load}+\sum_{j=1}^{N_i}\left[\frac{U_{i}^{2}}{X_{ij}}-\frac{U_{i}U_{j}}{X_{ij}}\cos{(\theta_{i}-\theta_{j})}\right]
    \end{aligned}
\end{align}
where $N_i$ denotes the number of DGs physically connected to DG~$i$; $P_{i,Load}$ and $Q_{i,Load}$ are active and reactive load demand at bus~$i$, and $U_i$ and $\theta_i$ are the bus voltage and the angle at bus $i$, respectively. {In the emergency condition, due to only maintaining the critical power supply, the line power flow exceeding to constraints is not considered. Moreover, we can select inductive output impedance $Z_{ij}$ and use virtual impedance~\cite{zhang2016distributed} in the control loop to cope with unknown equivalent impedance of power lines. Thus, we consider to control the power output of DGs to balance the power demand by a widely used droop-based power control loop~\cite{bidram2014multiobjective} } 
\begin{align}
    \begin{aligned}
        \omega_{i} & = \omega_{ni} - m_{Pi}P_{i} \\
	    U_{i} & = U_{ni}-n_{Qi}Q_{i}
    \end{aligned}\label{eq:mg_droop}
\end{align}
where $\omega_{ni},U_{ni}$ are set points of primary frequency and voltage control;  $\omega_{i},U_{i}$ are angular frequency and voltage magnitude of the $i_{\mathrm{th}}$ DG; $P_{i},Q_{i}$ are respectively active and reactive power outputs of the $i_{\mathrm{th}}$ DG; $m_{Pi},n_{Qi}$ are droop coefficients and are selected based on the active and reactive power ratings~\cite{bidram2014multiobjective}. As depicted in \cref{fig:cps_mg}, the primary controller consists of power control of \cref{eq:mg_droop} and inner control~\cite{ge2020extended}, through which the frequency and voltage deviations from the reference cannot be eliminated without effectively adjusting set-points. Hence, the secondary control is employed to achieve frequency regulation, accurate active power sharing, and voltage regulation, i.e.,
\begin{align}
	& \lim_{t\rightarrow\infty}\omega_{i}=\omega_{\mathrm{ref}},\quad\lim_{t\rightarrow\infty}\Bigg|\frac{P_{i}}{P_{\max,i}}-\frac{P_{j}}{P_{\max,j}}\Bigg|=0 \label{eq:obj_frequency} \\
	& \lim_{t\rightarrow\infty}U_{i}=U_{\mathrm{ref}}\label{eq:obj_voltage}
\end{align}
where $P_{\max,i}$ denotes the active power ratings of the $i_{\mathrm{th}}$ DG, and the second item of \cref{eq:obj_frequency} is equivalent to $\lim_{t\rightarrow\infty}|m_{Pi}P_{i}-m_{Pj}P_{j}|=0$ by appropriately setting $m_{Pi}P_{\max,i}=m_{Pj}P_{\max,j}$. \cref{eq:obj_voltage} only focuses on voltage regulation because voltage quality is prioritised for the critical supply in the emergency response, and voltage regulation and reactive power sharing cannot be reached simultaneously due to impedance effects except for a perfectly configuration~\cite{zuo2016distributed}.

\subsection{Post-Event Response under Pre-Scheduled Wireless Network}

\begin{algorithm}[!ht]
    \label{alg:2}
    \LinesNumbered
	\caption{Distributed Control Framework of Emergency MGs}
	\KwIn{$\omega_{ref},U_{ref}$:~reference~values~of~angular~frequency~and~voltage~magnitude; $K_{\omega{i}},K_{Pi},K_{Ui}$:~control~gains~of~angular~frequency,~active~power~sharing~and~voltage~magnitude; $\omega_{i}$:~local~angular~frequency; $m_{Pi}P_{i}$:~local~active~power~ratio; $U_{i}$:~local~voltage~magnitude; $m_{Pj}P_{j}$:~neighbouring~active~power~ratiod;}
	\KwOut{$\omega_{ni}$:~set~points~of~primary~frequency~control; $U_{ni}$:~set~points~of~primary~voltage~control;}
	\BlankLine
	\For{$i\in \mathcal{V}_{\mu}$, every $T_{s}^{\mu}$}{
	    update input variables (including delayed neighbouring information)\;
	    update set points in the primary control by \cref{eq:input}\;
	    send updated $m_{Pi}P_{i}$ to neighbours\;
	}
\end{algorithm}

The control system of post-event response suffers from neighbouring communication delay because the wireless C2C network has limited reserved bandwidth and each controller hardware has limited transmit power, as analysed in \cref{sec:cyber_schedule}. Therefore, we design a distributed control framework with sampling interval $T_{s}^{\mu}\geq\tau_{\max}^{\mu},\mu\in\Phi$. The existing transmission delay, as analysed in \cref{rem:delay} restricts the selection of sampling interval, i.e., $T_{s}^{\mu}<\tau_{\max}^{\mu},\mu\in\Phi$ could lead to the cumulative delay among the communication network, which makes the system suffer from unbounded time-varying delay, thereby raising the difficulty of controller design. Owing to the relatively low-frequency time-triggered control framework, for each emergency post-event MG, we model the dynamics of the system \cref{eq:mg_droop} in a discrete manner:
\begin{align}
    \begin{aligned}
	    \omega_{ni}(k+1) &=\omega_{i}(k+1)+m_{Pi}P_{i}(k+1)\\
	    &= \omega_{i}(k)+ u_{\omega i}(k) + m_{Pi}P_{i}(k) +u_{Pi}(k)
    \end{aligned}\\
    \begin{aligned}
        U_{ni}(k+1) &=U_{i}(k+1)+n_{Qi}Q_{i}(k+1)\\
	    &= U_{i}(k)+ u_{Ui}(k) + n_{Qi}Q_{i}(k) +u_{Qi}(k)
	\end{aligned}
\end{align}
Owing to the analysis for objective \eqref{eq:obj_voltage}, reactive power sharing control is omitted. Then, we obtain the equivalent matrix-form discrete model
\begin{align}   
    \left\{\begin{aligned}
        \bm{x}_{\omega}(k+1)=\bm{x}_{\omega}(k)+\bm{u}_{\omega}(k)\\
        \bm{x}_{P}(k+1)=\bm{x}_{P}(k)+\bm{u}_{P}(k)\\
        \bm{x}_{U}(k+1)=\bm{x}_{U}(k)+\bm{u}_{U}(k)
    \end{aligned}\right.\label{eq:discrete_model}
\end{align}
where $\forall{\mu\in\Phi}$,
\begin{align*}
    &\bm{x}_{\omega}=[x_{\omega{i}}]=[\omega_{i}]\in\mathbb{R}^{\mathcal{V}_{\mu}}, &\bm{u}_{\omega}=[u_{\omega{i}}]\in\mathbb{R}^{\mathcal{V}_{\mu}} \\
    &\bm{x}_{P}=[x_{Pi}]=[m_{Pi}P_{i}]\in\mathbb{R}^{\mathcal{V}_{\mu}}, &\bm{u}_{P}=[u_{Pi}]\in\mathbb{R}^{\mathcal{V}_{\mu}} \\
    &\bm{x}_{U}=[x_{Ui}]=[U_{i}]\in\mathbb{R}^{\mathcal{V}_{\mu}}, &\bm{u}_{U}=[u_{Ui}]\in\mathbb{R}^{\mathcal{V}_{\mu}}
\end{align*}

The distributed averaging proportional integral~(DAPI) discrete controller is formulated inspired by extensively studied consensus protocol~\cite{tian2008consensus}:
\begin{align}
\left\{\begin{aligned}
    & u_{\omega i}(k) = K_{\omega{i}}\big(x_{\omega,ref}-x_{\omega i}(k)\big)\\
    & u_{Pi}(k) = K_{Pi}\sum_{(i,j)\in\mathcal{E}_{\mu,i}}a_{ij}\big(x_{Pj}(k-1)-x_{Pi}(k)\big)\\
    & u_{Ui}(k) = K_{Ui}\big(x_{U,ref}-x_{Ui}(k)\big)
    \end{aligned}\right.\label{eq:input}
\end{align}
where $K_{\omega{i}},K_{Pi},K_{Ui}>0$ are the designed control gains. Thanks to the design of $T_{s}\geq\tau_{\max}$, one time step of delay caused by emergency wireless network is induced in \cref{eq:input}. 

\begin{thm}\label{thm:1}
    For emergency MGs controlled under Algorithm~\ref{alg:2} and \cref{eq:input} with a distributed C2C communication structure modelled by an undirected graph $\mathcal{G}$, the distributed frequency regulation, active power sharing and voltage regulation can be achieved as Eqs. \eqref{eq:obj_frequency} and \eqref{eq:obj_voltage} asymptotically if the conditions 
    \begin{subequations}
        \begin{align}
            & 0<K_{\omega{i}},K_{Ui}<2 \label{eq:thm_freq_vol}\\
            & 0<|\mathcal{E}_{\mu,i}|K_{Pi}<1 \label{eq:thm_p}
        \end{align}
    \end{subequations}
    are satisfied.
\end{thm}
\begin{proof}
    See \ref{sec:appendix1}.
\end{proof}

\begin{rem}
    \label{rem:1}
    The design principles Eqs.~\eqref{eq:thm_freq_vol} and \eqref{eq:thm_p} in \cref{thm:1} are delay independent, owing to that bounded communication delay does not affect the selection of control gains in consensus problems~\cite{liu2010consensus,munz2010delay}. Such bounded communication delay in the MG control of wireless-based post-event response derives from $T_{s}^{\mu}\geq\tau_{\max}^{\mu}$ in Algorithm~\ref{alg:2}. Larger $T_{s}^{\mu}$ may slower consensus, while $T_{s}^{\mu}<\tau_{\max}^{\mu}$ will lead to cumulative delays among wireless communications. Such time-varying and unbounded communication delays could lead to complicated design and dynamics~\cite{liu2010consensus}.
\end{rem}
\begin{rem}
    \label{rem:2}
    From \cref{eq:input}, only active power sharing, which balances the reserved power among dispatchable resources, requires cyber-layer wireless communication. In other words, frequency and voltage regulation only requires localised measurement. This means, before backup wireless communication is absolutely initialised, decentralised controlled MG systems can be stabilised, which is significant for emergency wireless formation. 
\end{rem}

\begin{figure}[!htb]
	\centering
	\includegraphics[width=0.8\textwidth]{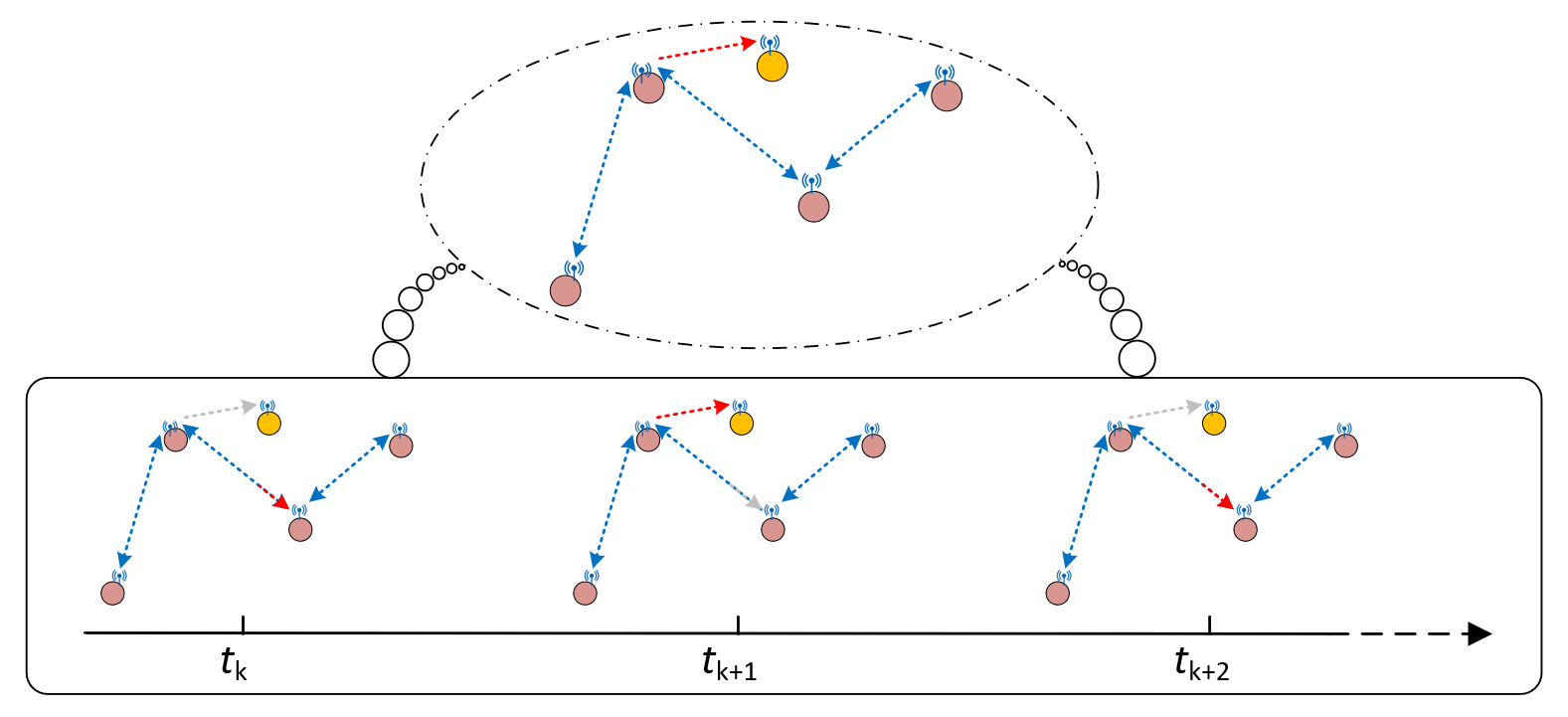}
	\caption{Cyber solution for mobile resources.}
	\label{fig:cyber_mobile}
\end{figure}
Mobile resources, e.g., portable energy storage and dispatchable electrified transportation have potentials to provide emergency response to extreme conditions~\cite{teng2016benefits,he2021utility}, and mobility requires the plug-and-play operation of MGs. Although plug-and-play operation has been widely investigated from a physical-layer perspective~\cite{ge2021event,ge2020resilient,ge2020extended}, it has not been investigated from a cyber-layer especially under extreme conditions. Due to limited communication bandwidth, just one-way communication is designed, i.e., mobile DGs only receive the information from networked DGs that have pre-designed in the cyber-layer network. One pre-designed and networked DG alternatively share information with one of previous fixed neighbours and one mobile DG, as shown in \cref{fig:cyber_mobile}, leading to second-order delay in the second item in \cref{eq:input}. The stability remains guaranteed by the analysis in \cref{rem:1} still using \cref{thm:1}. It is worth noting that the cyber solution of such plug-and-play operation cannot handle massive mobile resources, which is reasonable because most mobile devices have been dynamically scheduled, only limited plug-and-play operation needs to be mitigated in the post-event period.

{
\begin{rem}
    The proposed framework can cope with the increasing DGs and MGs in both normal condition and extreme condition. In the normal condition, the increasing DGs can be efficiently coordinated and regulated by base stations through appropriate bandwidth allocation algorithms. In the extreme condition, the increase in MG number will not affect the intra-MG communication network as a connected subgraph because the inter-MG interference/collision between wireless links is limited. In other words, the communication network of the intra-regional MG can reuse bandwidth resources with other regional MGs. Inside the MG, the increasing DG numbers and distances lead to increasing time delays. Although it will not affect the control gain by \cref{thm:1} to guarantee the stability, the sampling interval increases as the time delay, thus decreasing the convergence rate.
\end{rem}
}

\section{Results}\label{sec:results}
\begin{figure}[!ht]
	\centering
	\includegraphics[width=1.0\textwidth]{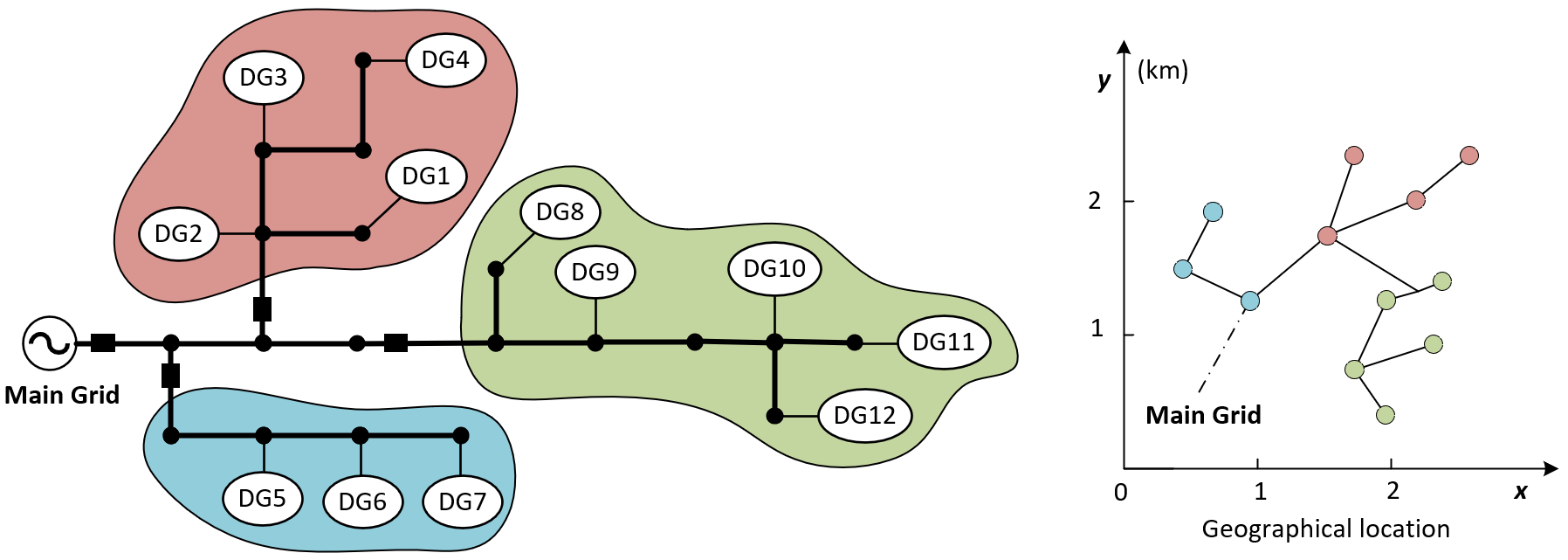}
	\caption{Diagram of the tested topology.}
	\label{fig:result_diagram}
\end{figure}

In this section, the centralised-to-decentralised cyber-physical cooperative response to enable the critical power supply is verified through the power network detailed in \cref{fig:result_diagram}, where three emergency MGs, naturally clustered by geographical locations are available to maintain critical power supply.

\subsection{Results of Dynamic Cyber-Layer Pre-Event Scheduling}

The parameters of the cyber layer are detailed in \ref{sec:appendix2}~\cite{kong2020multicell,termehchi2021joint}. Under the scenario with all DGs being dispatchable (Scenario 1), using Algorithm~\ref{alg:1}, the cyber layer of emergency MGs is scheduled by
\begin{align*}
    \mathcal{A} = \begin{bNiceMatrix}[first-row,first-col,code-for-first-row = \mathbf{\arabic{jCol}},code-for-first-col = \mathbf{\arabic{iRow}}]
    \CodeBefore
        \rectanglecolor[HTML]{D9958F}{1-1}{4-4}
        \rectanglecolor[HTML]{92CDDC}{5-5}{7-7}
        \rectanglecolor[HTML]{C4D6A0}{8-8}{12-12}
    \Body
        & 	& 	& 	& 	& 	& 	& 	& 	& 	& 	& 	& \\	
        & 0	& 1	& 1	& 0	& 	& 	& 	& 	& 	& 	& 	& 	\\
        & 1	& 0	& 0	& 0	& 	& 	& 	& 	& 	& 	& 	& 	\\
        & 1	& 0	& 0	& 1	& 	& 	& 	& 	& 	& 	& 	& 	\\
        & 0	& 0	& 1	& 0	& 	& 	& 	& 	& 	& 	& 	& 	\\
        & 	& 	& 	& 	& 0	& 1	& 0	& 	& 	& 	& 	& 	\\
        & 	& 	& 	& 	& 1	& 0	& 1	& 	& 	& 	& 	& 	\\
        & 	& 	& 	& 	& 0	& 1	& 0	& 	& 	& 	& 	& 	\\
        & 	& 	& 	& 	& 	& 	& 	& 0	& 1	& 0	& 1	& 0	\\
        & 	& 	& 	& 	& 	& 	& 	& 1	& 0	& 0	& 1	& 0	\\
        & 	& 	& 	& 	& 	& 	& 	& 0	& 0	& 0	& 0	& 1	\\
        & 	& 	& 	& 	& 	& 	& 	& 1	& 1	& 0	& 0	& 0	\\
        & 	& 	& 	& 	& 	& 	& 	& 0	& 0	& 1	& 0	& 0	
    \end{bNiceMatrix}
\end{align*}

From the adjacent matrix $\mathcal{A}$, we can find all three subgraphs are optimised to be connected as the emergency wireless network serving power supply locally. The minimised transmit delays $\bm{\tau}_{\max}=[87.5\ 55.4\ 44.6]^{T}(\mathrm{ms})$, hence the sampling intervals in the localised controllers of DGs are set as $\bm{T}_{s}=[100\ 60\ 50]^{T}(\mathrm{ms})$.

If DG~$2$, DG~$9$ and DG~$11$ are randomly out-of-service or non-dispatchable (Scenario 2), the corresponding cyber-layer scheduling result is
\begin{align*}
    \mathcal{A} = \begin{bNiceMatrix}[first-row,first-col,code-for-first-row = \mathbf{\arabic{jCol}},code-for-first-col = \mathbf{\arabic{iRow}}]
    \CodeBefore
        \rectanglecolor[HTML]{D9958F}{1-1}{4-4}
        \rectanglecolor[HTML]{92CDDC}{5-5}{7-7}
        \rectanglecolor[HTML]{C4D6A0}{8-8}{12-12}
    \Body
        & 	& 	& 	& 	& 	& 	& 	& 	& 	& 	& 	& \\	
        & 0	& -	& 1	& 0	& 	& 	& 	& 	& 	& 	& 	& 	\\
        & -	& -	& -	& -	& 	& 	& 	& 	& 	& 	& 	& 	\\
        & 1	& -	& 0	& 1	& 	& 	& 	& 	& 	& 	& 	& 	\\
        & 0	& -	& 1	& 0	& 	& 	& 	& 	& 	& 	& 	& 	\\
        & 	& 	& 	& 	& 0	& 1	& 0	& 	& 	& 	& 	& 	\\
        & 	& 	& 	& 	& 1	& 0	& 1	& 	& 	& 	& 	& 	\\
        & 	& 	& 	& 	& 0	& 1	& 0	& 	& 	& 	& 	& 	\\
        & 	& 	& 	& 	& 	& 	& 	& 0	& -	& 1	& -	& 0	\\
        & 	& 	& 	& 	& 	& 	& 	& -	& -	& -	& -	& -	\\
        & 	& 	& 	& 	& 	& 	& 	& 1	& -	& 0	& -	& 1	\\
        & 	& 	& 	& 	& 	& 	& 	& -	& -	& -	& -	& -	\\
        & 	& 	& 	& 	& 	& 	& 	& 0	& -	& 1	& -	& 0	
    \end{bNiceMatrix}
\end{align*}
with the minimised transmit delays $\bm{\tau}_{\max}=[63.3\ 55.4\ 174]^{T}(\mathrm{ms})$. Although the number of non-dispatchable DGs declines in the $3_{rd}$ MG compared to Scenario 1, the delay increases due to longer transmitting distances.

\subsection{Results of Post-Event Response}
\label{subsec:simulation2}
Take Scenario 1 as an example, the post-event response algorithm is verified. The performance of the designed DAPI discrete controller is evaluated with different gains, and mobile resources are also discussed further under plug-and-play operations.

\subsubsection{Response to Blackout}
\begin{figure}[!ht]
\centering
\includegraphics[width=1.0\textwidth]{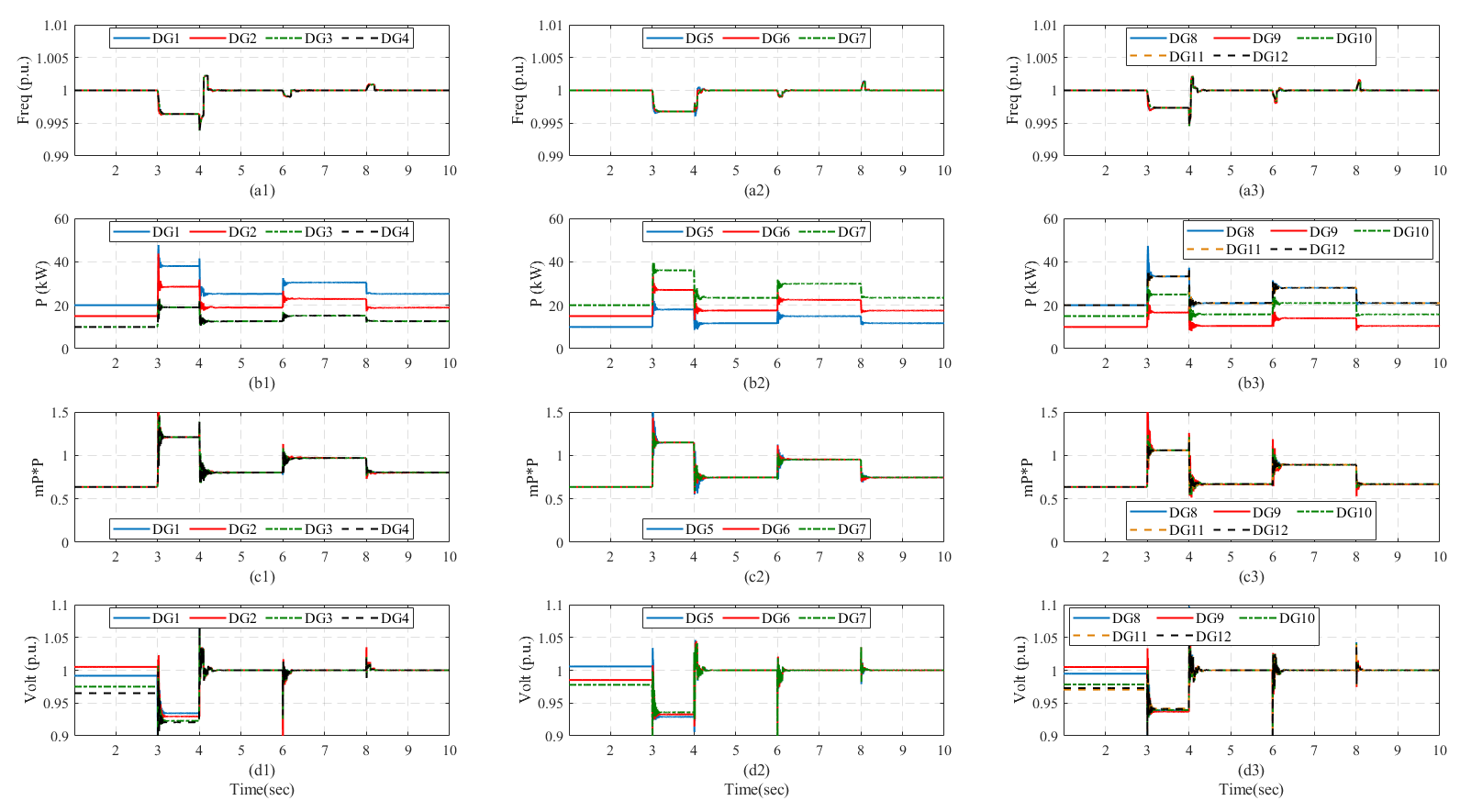}
\caption{Control performance with reasonable $\bm{K}_{\omega},\bm{K}_{P},\bm{K}_{U}$.}
\label{fig:bench}
\end{figure}

The control performance of the proposed C2C distributed control in response to blackouts is shown in~\cref{fig:bench}. After the occurrence of blackouts at $t=3\ \mathrm{seconds}$, owing to the grid-forming techniques, DGs maintain the critical power supply by emergency MGs in terms of geographical locations using primary control, which leads to the control deviation. Then, the secondary control is activated at $t=4\ \mathrm{seconds}$, when D2D-communication-based wireless network is completely initialised and only critical load demand is supplied. Followed by \cref{thm:1}, the stability of MGs is guaranteed, and the control objectives of Eqs. \eqref{eq:obj_frequency} and \eqref{eq:obj_voltage} can be reached, though the load demand changes at $t=6,8\ \mathrm{seconds}$.

\begin{figure}[!ht]
\centering
\includegraphics[width=1.0\textwidth]{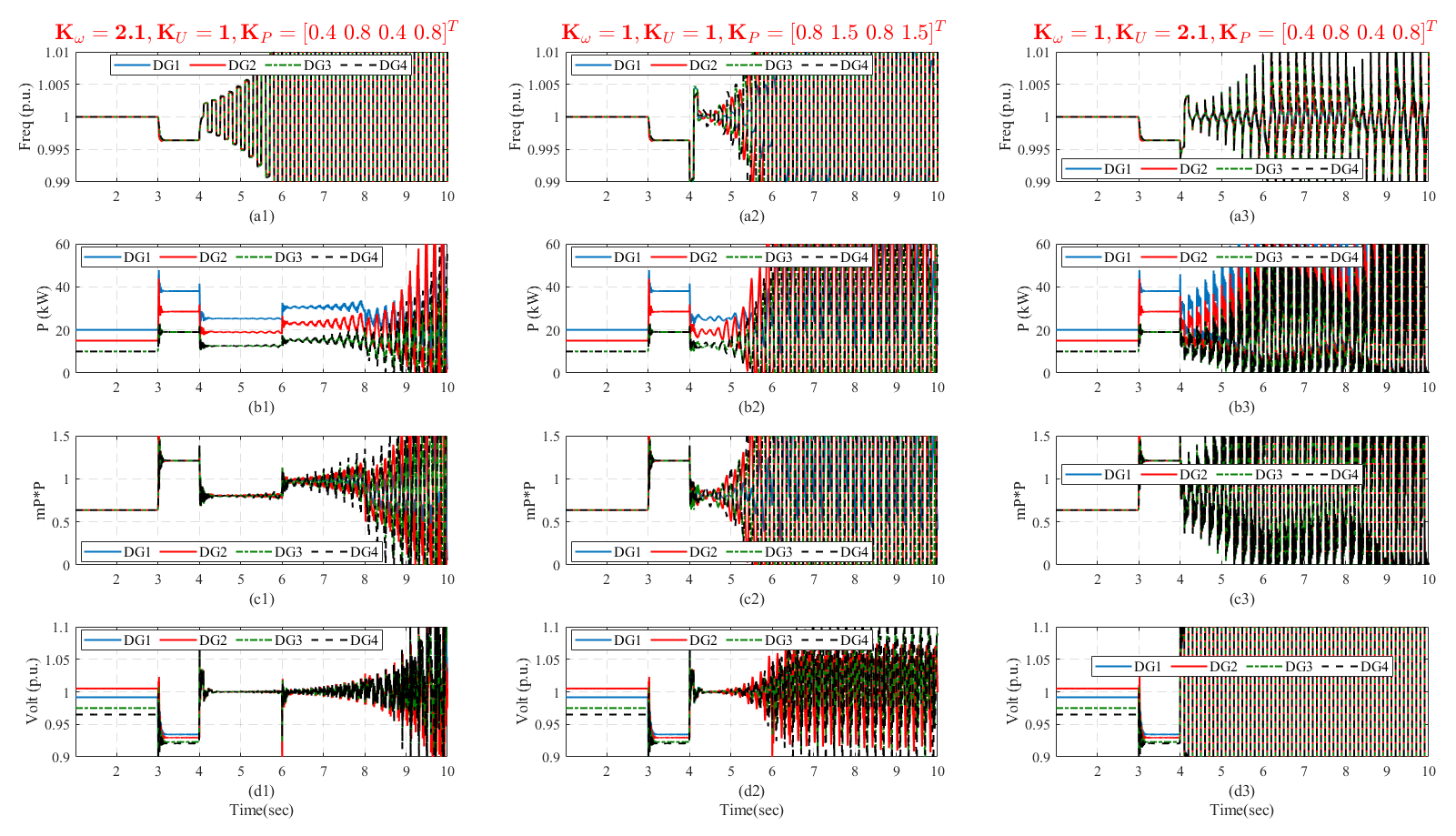}
\caption{Control performance with unreasonable $\bm{K}_{\omega},\bm{K}_{P},\bm{K}_{U}$.}
\label{fig:comparison}
\end{figure}
In order to verify the parameter design using \cref{thm:1}, we set the control gains different from that in the benchmark \cref{fig:bench}, i.e., $\bm{K}_{\omega} = \bm{1},\bm{K}_{U} = \bm{1},\bm{K}_{P}=[0.4\ 0.8\ 0.4\ 0.8]^{T}$ of the $1_{st}$ MG that satisfy \cref{thm:1}. From \cref{fig:comparison}, it is clear that the control performance degrades in all scenarios. More specifically, $\bm{K}_{\omega}=\bm{2.1}, \bm{K}_{U}=\bm{2.1}$,  and $\bm{K}_{P}=[0.8\ 1.5\ 0.8\ 1.5]^{T}$, exceeding to the boundaries of criteria Eqs. \eqref{eq:thm_freq_vol} and \eqref{eq:thm_p} respectively, lead to the divergency of frequency, voltage and active power sharing.

\subsubsection{Response to Mobile Resources}
\begin{figure}[!htb]
\centering
\includegraphics[width=0.5\textwidth]{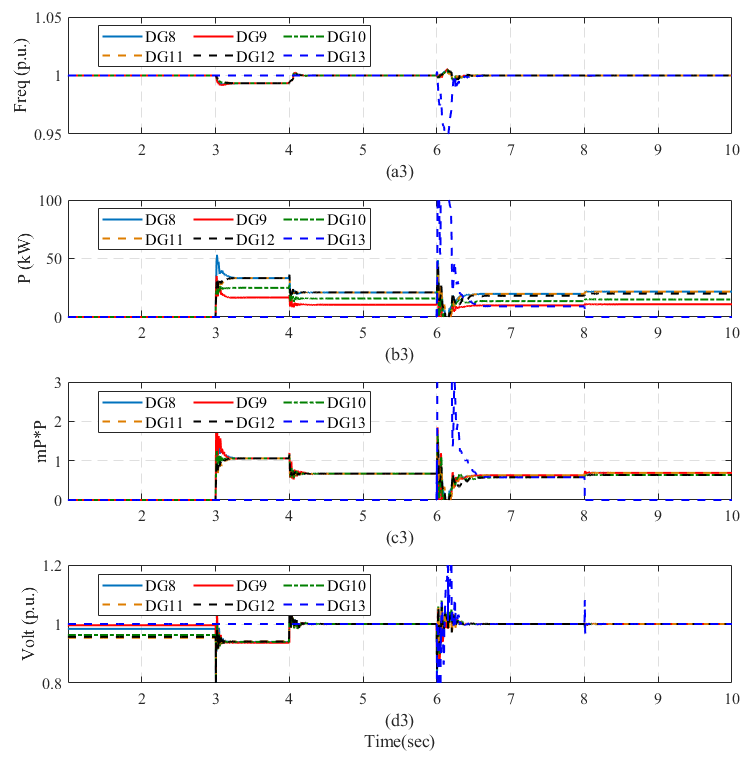}
\caption{Control performance with mobile resources.}
\label{fig:result_mobile}
\end{figure}
The scheduled cyber-layer wireless enables mobile resources providing emergency power supply, which is illustrated by \cref{fig:result_mobile}. The $3_{rd}$ MG is islanded at $t=3\ \mathrm{seconds}$ and the proposed C2C-enabled cyber-physical control strategy is activated at $t=4\ \mathrm{seconds}$. During $6\leq t\leq8\ \mathrm{seconds}$, the mobile resource, i.e., DG $13$, is plugged in the MG, and frequency, voltage and active power sharing remains controlled as Eqs.~\eqref{eq:obj_frequency},\eqref{eq:obj_voltage} though transient dynamics exist at the stage of emergency response. After disconnecting DG $13$ from the MG at $t=8\ \mathrm{seconds}$, the cyber-physical post-event response remains effective.

\subsection{Comparison of Post-Event Response Performance}
\begin{figure}[!ht]
\centering
\includegraphics[width=1.0\textwidth]{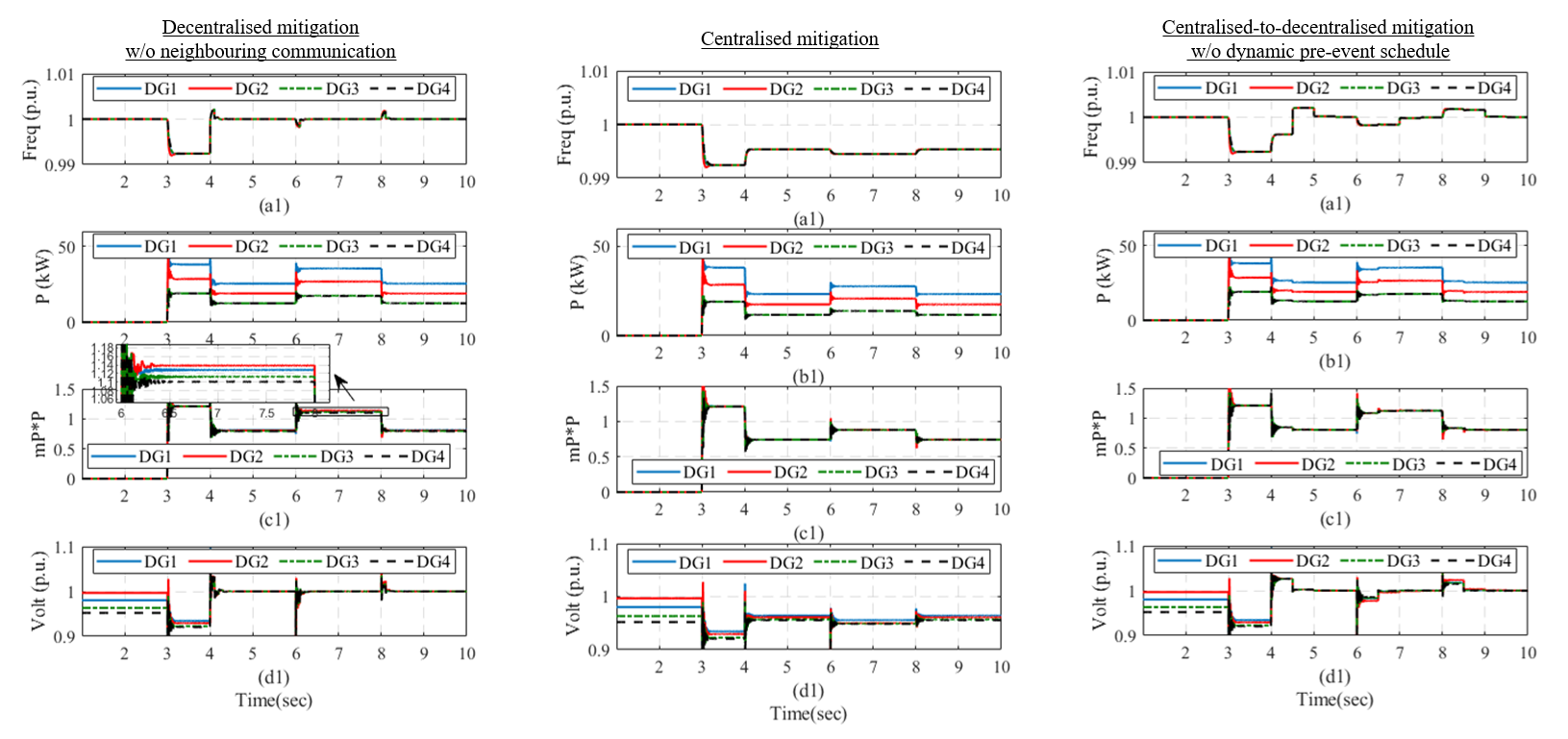}
\caption{Comparisons of post-event response performance.}
\label{fig:comparison2}
\end{figure}
As shown in~\cref{fig:comparison2}, the post-event response performance of the $1st$ MG under proposed framework is compared with other solutions, including  decentralised strategy without neighbouring communication, centralised strategy and centralised-to-decentralised mitigation without dynamic pre-event schedule. The decentralised mitigation without neighbouring communication~(first column) regulates the frequency and voltage to their references, while the power sharing is not accurately guaranteed, see the third row. The centralised mitigation utilises the same bandwidth that is allocated in \cref{subsec:simulation2}. Due to the long distance of data flow in the centralised framework, the delays~(8.35 seconds) induced by the limited bandwidth lead to the uncontrollable period between two sampling intervals. If the centralised-to-decentralised mitigation without dynamic schedule strategy is applied, the scheduled delays should be conservative by the consideration of the worst case, leading to a large sampling interval. Such design degrades the post-event response performance as shown in the third column. All three solutions have not achieved the optimised response performance, compared to \cref{fig:bench}. Although the basic stability can be guaranteed by the grid-forming technique,  optimised voltage, frequency, and power sharing may be compromised under such emergency conditions.

\section{Conclusion}\label{sec:conclusion}
This paper proposes a centralised-to-decentralised framework to enhance the resilience of power supply in response to possible failures/blackouts caused by ACPEs. In the proposed resilient framework, the cyber-physical response plan is dynamically updated in the centralised controller of networked MGs under the normal operation, where cyber-layer C2C communication and physical-layer emergency MGs formation are pre-scheduled. Considering the possible damage of base stations, the backup communication employs dedicated wireless network to provide reliable services for real-time control. The inevitable delay derived from the backup bandwidth is then considered in the distributed control system design. At last, the whole pre-event scheduling process and post-event response performance are evaluated through the case studies.

\section*{Acknowledgements}
This work was supported by EPSRC under Grant EP/T021780/1 and by The Royal Society under Grant RGS/R1/211256.

\appendix
\section{Proof of Theorem~\ref{thm:1}}
\label{sec:appendix1}
Define $\bm{x}^{e}_{\omega}=\bm{x}_{\omega}-x_{\omega,ref}\mathbf{1}_{N},\bm{x}^{e}_{U}=\bm{x}_{U}-x_{U,ref}\mathbf{1}_{N}$, where $\mathbf{1}_{N}$ denotes a column vector with all elements being ones, and $\bm{K}_{\star}=\mathrm{diag}{\{K_{\star{i}}\}}\in\mathbb{R}^{V_{\mu}\times{V_{\mu}}},\star\in\{\omega,P,U\}$ the dynamics of the system \eqref{eq:discrete_model} can be expressed in a matrix form by identity matrix $\bm{I}$ of appropriate dimensions:
\begin{subequations}
	\begin{align}
	&\bm{x}^{e}_{\omega}(k+1)=(\bm{I}-\bm{K}_{\omega})\bm{x}^{e}_{\omega}(k) \label{eq:dynamic_freq} \\
	&\bm{x}_{P}(k+1)=\bm{x}_{P}(k)-\bm{K}_{P}\mathcal{D}_{\mu}\bm{x}_{P}(k)+\bm{K}_{P}\mathcal{A}_{\mu}\bm{x}_{P}(k-1) \label{eq:dynamic_power} \\
	&\bm{x}^{e}_{U}(k+1)=(\bm{I}-\bm{K}_{U})\bm{x}^{e}_{U}(k) \label{eq:dynamic_vol}
	\end{align}
\end{subequations}
where $\mathcal{D}_{\mu}=\mathrm{diag}\{d_{i}\}\in\mathbb{R}^{V_{\mu}\times{V_{\mu}}}, d_{i}=|\mathcal{E}_{\mu,i}|$ is the diagonalised out-degree matrix of $\mathcal{G}_{\mu}$.

Firstly, we give the proof of the stability of the frequency regulation. Taking $z$-transformation of the system~\eqref{eq:dynamic_freq}
\begin{align*}
	z\bm{X}_{\omega}^{e}(z)=(\bm{I}-\bm{K}_{\omega}){X}_{\omega}^{e}(z)
\end{align*}
the characteristic equation of which is
\begin{align}
	\mathrm{det}\big((z-1)\bm{I}+\bm{K}_{\omega}\big)=0
	\label{eq:det_freq}
\end{align}
Owing to the diagonalised form of $\bm{K}_{\omega}$, the characteristic equation~\eqref{eq:det_freq} is equivalent to $z-1+K_{\omega{i}}=0 \Longrightarrow z=1-K_{\omega{i}}$. By the stability criteria that the system \eqref{eq:dynamic_freq} is asymptotically stable if the roots of \cref{eq:det_freq} have modulus less than unity, the asymptotical stability of the system~\eqref{eq:dynamic_freq} is guaranteed if the following condition is satisfied:
\begin{align*}
	|z|=|1-K_{\omega{i}}|<1\Longrightarrow0<K_{\omega{i}}<2
\end{align*}
Similarly, the asymptotically stability of the system \eqref{eq:dynamic_vol} is guaranteed by $0<K_{Ui}<2$. Therefore, the proof with regard to \cref{eq:thm_freq_vol} is completed.

Then, we give the proof for the stability of the active power sharing. Taking $z$-transformation of the system~\eqref{eq:dynamic_power}
\begin{align}
	\begin{aligned}
		z\bm{X}_{P}(z)&=\bm{X}_{P}(z)-\bm{K}_{P}\mathcal{D}_{\mu}\bm{X}_{P}(z)+z^{-1}\bm{K}_{P}\mathcal{A}_{\mu}\bm{X}_{P}(z) \\
		&= \bm{X}_{P}(z)-\widetilde{\bm{L}}_{\mu}\bm{X}_{P}(z)
	\end{aligned}
\end{align}
where
\begin{align*}
	\widetilde{\bm{L}}_{\mu}(z) = \bm{K}_{P}\mathcal{D}_{\mu}-z^{-1}\bm{K}_{P}\mathcal{A}_{\mu}
	= \left\{\begin{aligned}
		& -K_{Pi}a_{ij}z^{-1}, && (i,j)\in\mathcal{E}_{\mu} \\
		& K_{Pi}d_{i}, && i\in\mathcal{V}_{\mu} \\
		& 0, && \mathrm{otherwise}
	\end{aligned}\right\}\in\mathbb{R}^{V_{\mu}\times V_{\mu}}
\end{align*}
It should be noted that $\widetilde{\bm{L}}_{\mu}(1)=\bm{L}_{\mu}$ denotes the Laplacian matrix of $\mathcal{G}_{\mu}$. Define $p(z)=\mathrm{det}\big((z-1)\bm{I}+\widetilde{\bm{L}}_{\mu}(z)\big)$, then the asymptotical stability of the system~\eqref{eq:dynamic_power} is guaranteed by all the zeros of $p(z)$ having modulus less than unity except for a zero at $z=1$~(see, e.g., \textit{Lemma 1} in \cite{tian2008consensus}).

Since graph $\mathcal{G}_{\mu}$ is undirected and connected, $0$ is one eigenvalue of $\bm{L}_{\mu}$ and $\mathrm{rank}(\bm{L}_{\mu})=V_{\mu}-1$~\cite{lin2005necessary}, thereby $p(1)=\mathrm{det}(\bm{L}_{\mu})=0$ showing that $z=1$ is indeed one of zeros.

Next, we prove that the zeros of $f(z)=\mathrm{det}\big(\bm{I}+\frac{\widetilde{\bm{L}}_{\mu}(z)}{z-1}\big)$ have modulus less than unity. It is achieved if the eigenvalue loci of $\frac{\widetilde{\bm{L}}_{\mu}(e^{j\omega})}{e^{j\omega}-1}$, i.e., $\lambda\big(\frac{\widetilde{\bm{L}}_{\mu}(e^{j\omega})}{e^{j\omega}-1}\big),\forall\omega\in[-\pi,\pi]$ does not enclose $(-1,j0)$ in terms of the fact $K_{Pi},a_{ij}>0$ based on general Nyquist stability criteria. Using Gerschgorin disk theorem, we have 
\begin{align*}
	\lambda\Bigg(\frac{\widetilde{\bm{L}}_{\mu}(e^{j\omega})}{e^{j\omega}-1}\Bigg)\in\bigcup_{i\in\mathcal{V}_{\mu}}\mathcal{S}_{i},\forall\omega\in[-\pi,\pi]
\end{align*}
\begin{align*}
	\mathcal{S}_{i}=&\Bigg\{s\in\mathbb{C}:\bigg|s-\frac{K_{Pi}d_{i}}{e^{j\omega}-1}\bigg|\leq\sum_{(i,j)\in\mathcal{E}_{\mu}}\bigg|\frac{K_{Pi}a_{ij}e^{-j\omega}}{e^{j\omega}-1}\bigg|\leq\bigg|\frac{K_{Pi}d_{i}}{e^{j\omega}-1}\bigg|\Bigg\}
\end{align*}
\begin{figure}[!hb]
	\centering
	\includegraphics[width=0.6\textwidth]{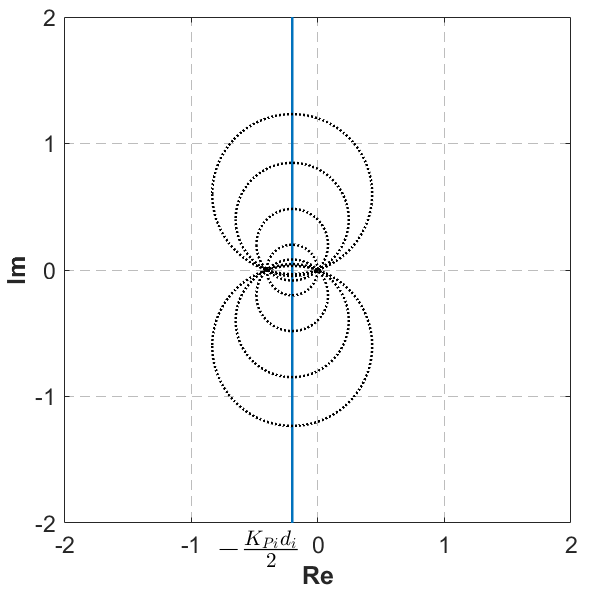}
	\caption{Nyquist plot of $F(jw)$.}
	\label{fig:nyquist}
\end{figure}
Define
\begin{align*}
	 F(j\omega)=\frac{K_{Pi}d_{i}}{e^{j\omega}-1}=-\frac{K_{Pi}d_{i}}{2}-j\frac{K_{Pi}d_{i}\cos{\frac{\omega}{2}}}{2\sin{\frac{\omega}{2}}}
\end{align*}
which is the center of $\mathcal{S}_{i}$ and its modulus is the radius. The trajectory of $F(jw)$, i.e. Nyquist plot, and the corresponding Gerschgorin disks are shown in \cref{fig:nyquist}, from which we find $\lambda\big(\frac{\widetilde{\bm{L}}_{\mu}(e^{j\omega})}{e^{j\omega}-1}\big),\forall\omega\in[-\pi,\pi]$ does not enclose $(-1,j0)$ as long as $(-1,j0)$ is outside $\mathcal{S}_{i}$, i.e., $(-1,j0)\notin\mathcal{S}_{i}$:
\begin{align*}
	|-1-F(j\omega)|^{2}-|F(j\omega)|^{2}=1-K_{Pi}d_{i}=1-|\mathcal{E}_{\mu,i}|K_{Pi}>0
\end{align*}
which is satisfied by \cref{eq:thm_p}.

Therefore, \cref{thm:1} is proved.

\section{Parameters of Dynamic Cyber-Layer Wireless Network Scheduling}
\label{sec:appendix2}
Distance Matrix (km):
\begin{align*}
   \begin{bNiceMatrix}[first-row,first-col,code-for-first-row = \mathbf{\arabic{jCol}},code-for-first-col = \mathbf{\arabic{iRow}}]
    \CodeBefore
        \rectanglecolor[HTML]{D9958F}{1-1}{4-4}
        \rectanglecolor[HTML]{92CDDC}{5-5}{7-7}
        \rectanglecolor[HTML]{C4D6A0}{8-8}{12-12}
    \Body
        & 	& 	& 	& 	& 	& 	& 	& 	& 	& 	& 	& \\	
        & 0	& 0.64	& 0.56	& 0.86	& 	& 	& 	& 	& 	& 	& 	& 	\\
        & 0.64	& 0	& 0.72	& 1.20	& 	& 	& 	& 	& 	& 	& 	& 	\\
        & 0.56	& 0.72	& 0	& 0.52	& 	& 	& 	& 	& 	& 	& 	& 	\\
        & 0.86	& 1.20	& 0.52	& 0	& 	& 	& 	& 	& 	& 	& 	& 	\\
        & 	& 	& 	& 	& 0	& 0.55	& 0.72	& 	& 	& 	& 	& 	\\
        & 	& 	& 	& 	& 0.55	& 0	& 0.48	& 	& 	& 	& 	& 	\\
        & 	& 	& 	& 	& 0.72	& 0.48	& 0	& 	& 	& 	& 	& 	\\
        & 	& 	& 	& 	& 	& 	& 	& 0	& 0.45	& 0.93	& 0.48	& 1.10	\\
        & 	& 	& 	& 	& 	& 	& 	& 0.45	& 0	& 0.58	& 0.48	& 0.86	\\
        & 	& 	& 	& 	& 	& 	& 	& 0.93	& 0.58	& 0	& 0.62	& 0.41	\\
        & 	& 	& 	& 	& 	& 	& 	& 0.48	& 0.48	& 0.62	& 0	& 0.64	\\
        & 	& 	& 	& 	& 	& 	& 	& 1.10	& 0.86	& 0.41	& 0.64	& 0	
    \end{bNiceMatrix}
\end{align*}

\begin{table}[!ht]
	\centering
	\caption{Parameters of Pre-event C2C Wireless Network Scheduling}
	\begin{tabular}{c|c}
		\hline\hline
		Parameter & Value \\
		\hline\hline
		sub-carrier bandwidth ($w$) & 25 kHz\\
		\hline
		number of sub-carriers ($L$) & 40 \\
		\hline
		maximum transmission power ($P_{i,\max}$) & 24 dBm \\
		\hline
		constant power ($P_{i,\mathrm{cst}}$) & 0.1 dBm \\
		\hline
		noise power ($\sigma^{2}$) & -62 dBm \\
		\hline
		packet size ($L_\mathrm{packet}$) & 32 bytes\\
		\hline
		pathloss exponent ($\alpha$) & 3 \\
		\hline
		loss factor ($h$) & 0.09 \\
		\hline\hline
	\end{tabular}
\end{table}

\bibliographystyle{elsarticle-num} 
\bibliography{cas-refs}

\end{document}